\documentclass[12pt]{JHEP3}

\usepackage{epsfig}

\def\eq{\begin{equation}}
\def\eeq{\end{equation}}
\def\eqa{\begin{eqnarray}}
\def\eeqa{\end{eqnarray}}
\def\ba{\begin{array}}
\def\ea{\end{array}}
\def\nn{\nonumber}

\def\bd{\begin{displaymath}}
\def\ed{\end{diplaymath}}
\def\cA{{\cal A}}
\def\cF{{\cal F}}
\def\cL{{\cal L}}
\def\cO{{\cal O}}
\def\cD{{\cal D}}
\def\g{\gamma}
\def\gl{\gamma_{\scriptscriptstyle L}}
\def\gr{\gamma_{\scriptscriptstyle R}}
\def\gf{\gamma_5}
\def\d{\delta}
\def\e{\eta}
\def\diag{\hbox{diag}}
\def\pl{\partial}
\def\hf{{1\over 2}}
\def\qt{{1\over 4}}

\def\ol#1{\overline{#1}}
\def\Dsl{\hbox{/\kern-.6700em\it D}} 
\def\dsl{\hbox{/\kern-.5300em$\partial$}}
\def\veps{\varepsilon}
\def\eps{\epsilon}
\def\ebar{\ol{\eta}}
\def\pbar{\ol{\psi}}
\def\zbar{\ol{\zeta}}
\def\xbar{\ol{\xi}}
\def\lbar{\ol{\lambda}}
\def\cbar{\ol{\chi}}
\def\abar{\ol{\alpha}}
\def\bbar{\ol{\beta}}
\def\cc{\hbox{c.c.}}
\def\ssr{{\scriptscriptstyle R}}

\def\pref#1{(\ref{#1})}

\def\Box{ {\,\lower 0.9pt\vbox{\hrule\hbox{\vrule height0.2cm \hskip 0.2cm \vrule height 0.2cm }\hrule}\,}}

\def\lsim{{\ \lower-1.2pt\vbox{\hbox{\rlap{$<$}\lower5pt\vbox{\hbox{$\sim$}}}}\ }}

\def\gsim{{\ \lower-1.2pt\vbox{\hbox{\rlap{$>$}\lower5pt\vbox{\hbox{$\sim$}}}}\ }}

\def\pref#1{(\ref{#1})}

\title{Low-Energy Brane-World Effective Actions
and Partial Supersymmetry Breaking}
\author{C.~P.~Burgess$^1$, E.~Filotas$^1$, M.~Klein$^2$ and F.~Quevedo$^3$
\\
$^1$ Physics Department, McGill University,  3600 University Street,\\
              Montr\'eal, Qu\'ebec, Canada, H3A 2T8.\\

$^2$ Stanford Linear Accelerator Center, Stanford University,\\
Stanford, CA 94309, USA.\\

$^3$ Centre for Mathematical Sciences, DAMTP,
               University of Cambridge,\\
               Cambridge CB3 0WA, UK.}

\abstract{As part of a programme for the general study of the
low-energy implications of supersymmetry breaking in brane-world
scenarios, we study the nonlinear realization of supersymmetry
which occurs when breaking $N=2$ to $N=1$ supergravity. We
consider three explicit realizations of this supersymmetry
breaking pattern, which correspond to breaking by one brane, by one
antibrane or by two (or more) parallel branes. We derive the
minimal field content, the effective action and supersymmetry
transformation rules for the resulting $N=1$ theory perturbatively
in powers of $\kappa = 1/M_{Planck}$. We show that the way the
massive gravitino and spin-1 fields assemble into $N=1$ multiplets
implies the existence of direct brane-brane contact interactions
at order $\cO(\kappa)$. This result is contrary to the
$\cO(\kappa^2)$ predicted by the sequestering scenario but in
agreement with recent work of Anisimov {\it et al}. Our low-energy
approach is model independent and is a first step towards
determining the low-energy implications of more realistic brane
models which completely break all supersymmetries.}

\keywords{supersymmetry breaking, D-branes}

\preprint{McGill-02/31\\SLAC-PUB-9511\\DAMTP-2002-120}

\begin{document}

\section{Introduction}
String theory is our most promising theory of physics at the
highest energies, and predicts that the world is supersymmetric at
a very fundamental level. Unfortunately, it does not yet seem to
unambiguously predict how this supersymmetry gets broken as it
must be in order to make contact with the low-energy world of
present-day experiments. An understanding of supersymmetry
breaking in string theory appears to be a crucial prerequisite for
bringing string theory into the mainstream of physical thought.

The discovery of D-branes \cite{DBranes} introduced a fundamentally
new mechanism for supersymmetry breaking in string theory, since
each D-brane typically breaks half of the supersymmetries of the
theory. This mechanism differs from others which are usually
considered when model building with supersymmetric field theories,
in that it evades \cite{PH} an old no-go theorem \cite{NoGo} which
was thought to rule out the possibility of partially breaking an
extended supersymmetry.

Having the branes themselves break supersymmetry is a particularly
appealing picture within a brane-world scenario, for several
reasons.
\begin{itemize}
\item
If the lowest-energy supersymmetry is broken on a brane which is
physically separated from the brane on which all observed
particles are trapped, then we have a natural realization of the
hidden-sector models which have been long recognized as being of
great phenomenological interest. Since the interbrane couplings
are generically of gravitational strength, the supersymmetry
breaking scale is naturally much smaller than the string scale,
potentially allowing a novel understanding of the origin of the
electroweak hierarchy \cite{BBIQ}.
\item
`Realistic' string models having low-energy spectra which contain
the known Standard Model fields have been constructed
\cite{Models}, for which all observed particles are trapped on a
brane and supersymmetry is broken purely by the configuration of
branes considered.
\item
A framework where supersymmetry partially breaks on distant
branes, with the news of this breaking reaching us only through a
bulk space which has more supersymmetry than does our brane, has
the potential of alleviating some of the naturalness problems
which usually plague supersymmetry breaking. Although the
consequences of the extra bulk supersymmetries have not yet been
explored for the flavor problem of supersymmetry breaking, it has
been argued to have the potential to help with the cosmological
constant problem \cite{BMQ}.
\end{itemize}

In addition to having a bulk sector which enjoys extended
supersymmetry, another very interesting brane-based mechanism has
been proposed for suppressing, within supersymmetric models, the
direct couplings between fields which live on different branes
which are separated from one another in the extra dimensions.
According to this mechanism, called {\it sequestering} \cite{RS0},
the combination of locality in the extra dimensions with $N=1$
supersymmetry in 4 dimensions suppresses the couplings between
fields which live on different branes. For instance, it is
proposed that the K\"ahler potential, $K_{\rm tot}$, for the total
4D low-energy theory describing a two-brane configuration, has the
form
\eq \label{E:sequester}
   \exp\left( - \; {\kappa^2 K_{\rm tot} \over 3} \right) =
   \exp\left( - \; {\kappa^2 K_{1} \over 3} \right) + \exp\left( - \;
{\kappa^2 K_{2} \over 3} \right), \eeq
where $K_1$ and $K_2$ are the K\"ahler functions describing the 4D
supergravity separately on each of the branes. Here $\kappa^2 = 8
\pi G = 1/M_p^2$ is the 4D gravitational coupling constant. If
true, this remarkable formula would strongly constrain the kinds
of direct couplings which could arise between brane fields,
allowing other induced couplings -- such as anomaly-mediated
interactions \cite{RS0,AM} -- to dominate.

The authors of ref.~\cite{ADGT} examined this sequestered form in
various specific string compactifications, and found that it was
not generically satisfied in the string vacua examined. The
failure of this form was traced by these authors either to the
exchange of various bulk supergravity modes, or to the warping of
the compact dimensions.

In this paper we perform a complementary study of sequestering, by
directly examining the implications of  both the unbroken and the
nonlinearly-realized supersymmetries for the low-energy 4D
effective action for {\it any} brane model which spontaneously
breaks $N=2$ supersymmetry to $N=1$ supersymmetry. We show that
the way the massive gravitino and spin-1 fields assemble into
$N=1$ supersymmetry multiplets implies the existence of direct
brane-brane contact interactions already at $\cO(\kappa)$, in
conflict with the $\cO(\kappa^2)$ size which would be predicted by
eq.~\pref{E:sequester}. In this way we show that the failure of
the sequestered form is quite robust, and cannot hold in any brane
configuration which breaks $N=2$ supersymmetry down to $N=1$. We
believe it to be an open question whether sequestering can occur
in realistic brane-world models, and we discuss some evidence that
it might require additional supersymmetries in the concluding
section.

More generally, our investigation can be seen as a first step
towards the study of the low-energy limit of more realistic models
for which supersymmetry is completely broken by brane
configurations situated within bulk spacetimes enjoying extended
supersymmetries. Indeed this is our prime motivation for embarking
on this programme, and we consider it as the natural next step in
this direction. In particular, if brane supersymmetry breaking is
to play a role in solving naturalness problems at low energies, it
should be possible to understand how this is done purely within
the effective theory below the compactification scale, within
which all the information about brane separations has already been
integrated out. For this reason we cast our discussion completely
in terms of how supersymmetry is realized within the low-energy
four-dimensional theory. (For recent discussions of supersymmetry
breaking in the brane-world scenario from a more microscopic 
perspective see, for instance, \cite{DM}.)

The questions which such a study of low-energy supersymmetry
breaking could hope to address include: finding the effective
scale of mass splittings within supermultiplets, as compared to
the string and compactification scales; the structure of soft- and
hard-breaking interactions, with a comparison with the standard
$F$- and $D$-term breaking in more standard supergravity-mediated
models; the consequences of extended supersymmetries for gauge-
and anomaly-mediated supersymmetry breaking; the relative
suppression of the remnant value of the cosmological constant as
compared to the scale of multiplet splitting. (First steps along
some of these lines have been made for 5D theories, such as by
establishing the connection between $F$-term supersymmetry
breaking in the low-energy theory and Scherk-Schwarz supersymmetry
breaking in microscopic brane models \cite{MP}.) A proper study of
all of these issues lies at the core of any understanding of the
phenomenological implications of D-brane models.

Our presentation is organized as follows. First we describe how
the low-energy states associated with the gravitini are organized
in a model which breaks $N=2$ supersymmetry down to $N=1$ in four
spacetime dimensions. The generalization of this result to an
antibrane is also briefly described. We then extend this analysis
to the case where more than two branes participate in breaking the
one supersymmetry, and so implying that the would-be Goldstone
modes are a mixture of fields which arise on the two branes. In
this case we show how the assembly of states into supermultiplets
of the unbroken $N=1$ supersymmetry require the existence of
direct brane couplings at $\cO(\kappa)$. Our conclusions are
briefly summarized in the final section. A warning about notation:
we present our results using the 4-component spinor notation which
is commonly used outside of the supersymmetry community. However,
for the benefit of the reader we also present in an appendix a
dictionary which makes the explicit connection with two-component
Weyl-spinor notation. Further technical details on transformation
rules for massive spin $3/2$ and spin $1$ are presented in a second
appendix.

\section{The Single Brane}
Imagine now any configuration of branes within a
higher-dimensional spacetime that is compactified to flat four
dimensions, for which a 4D $N=2$ supersymmetry is unbroken.
Suppose also that the brane configuration breaks this $N=2$
supersymmetry, leaving an $N=1$ supersymmetry unbroken
\cite{susybranebreaking}.

Our purpose in this section is to describe the universal part of
the low-energy limit of such a configuration, focussing on those
very low energy degrees of freedom whose mass is well below the
model's compactification scale $E \ll 1/r$. The goal is to
identify general features of the 4D effective theory defined after
all of the extra-dimensional physics is integrated out, to see how
the supersymmetry-breaking pattern is realized in the low-energy
theory \cite{PSSB,BA}. Our analysis closely follows the pioneering
work of ref.~\cite{BA}, although we do find a few minor
corrections to some of the results found in that paper.

Following ref.~\cite{BA}, we obtain the low-energy theory by using
the following two properties which the spectrum of any such brane
construction enjoys.
\begin{itemize}
\item
First, the spectrum necessarily involves two spin-3/2 gravitini.
One of these is massless (for the unbroken supersymmetry) while
the other is massive (for the broken supersymmetry). Although
massive, this second gravitino is nevertheless within the
low-energy theory because its mass can be well below the
compactification scale, $1/r$.
\item
Second, the unbroken supersymmetry organizes these two gravitini
into $N=1$ supermultiplets, with the massless gravitino being
paired with the graviton and the massive gravitino combining with
two massive spin-1 and one massive spin-1/2 particle into a
massive $N=1$ spin-3/2 multiplet.
\end{itemize}
The strategy is to identify the brane and bulk degrees of freedom
by writing down the lagrangian for these two gravitino multiplets,
and then {\it unHiggsing} the massive spin-1 and spin-3/2 fields
\cite{BA} to identify the would-be Goldstone modes which are eaten
to produce these massive states.

\subsection{The Massive $N=1$ Spin 3/2 Multiplet}
Our starting point is the free lagrangian for a massive spin-3/2
multiplet in $N=1$ global supersymmetry. This multiplet must
contain two massive spin-1 ($\cA_m = (A_m + i B_m)/\sqrt{2}$) and
one massive spin-1/2 particle ($\zeta$) as well as the massive
spin-3/2 field ($\eta_m$). Its free action may therefore be
written:
\eqa \label{E:Lkin}
    \cL_{\rm kin} &=& - \, {i \over 2}
    \epsilon^{mnpq} \ol{\eta}_m \gf \g_n \pl_p \eta_q - {1\over 2}
    \cA^{mn} \cA_{mn}^* - \hf \, \ol{\zeta} \dsl \zeta \nn\\
    &&  \qquad - \, {m \over 2} \; \ol{\eta}_m \g^{mn} \eta_n - {m^2} \,
        \cA^{m} \cA_{m}^* - {m\over 2} \; \ol{\zeta} \zeta,
\eeqa
where $\cA_{mn} = \pl_m \cA_n - \pl_n \cA_m$ satisfies
$\epsilon^{mnpq} \pl_n \cA_{pq} = 0$. In our conventions the
fermionic fields, $\zeta$ and $\eta_m$, are chosen to be Majorana
spinors. The freedom to rescale the fields has been used to put
all kinetic terms into canonical form.

As is easily verified, this action enjoys an invariance with
respect to the following global supersymmetry transformations
\cite{FvN}:
\eqa \label{E:SusyTransf}
\d \cA_m &=& \sqrt2 (\ol\eta_m \gl
\veps) + {2 \over \sqrt6}
 (\zbar \g_m \gl \veps) - {2\over \sqrt{6} m}
\pl_m \Bigl( \zbar \gl \veps \Bigr), \\
\gl \d \zeta &=& {1 \over \sqrt6} \cA^*_{mn} \; \g^{mn} \gl \veps
-\, {2 m \over \sqrt6} \, \cA_m \, \g^m \gr \veps, \\
\gl \d \eta_m &=& {1 \over m} \pl_m \left({\sqrt 2 \over 6}
\cA_{ab}^* \g^{ab} \gl \veps + {2 \sqrt{2} m\over 3}
 \cA_a \g^a \gr \veps \right) -\, {2 \sqrt2 \over 3}
 \cA_{mn} \g^n \gr \veps \nn\\
&& + {i \sqrt2 \over 6} \eps_{mnab} \cA^{ab} \g^n \gr \veps -\,
{\sqrt{2} m\over 3} \cA^{n*} \g_{mn} \gl \veps + {2\sqrt{2} m\over3}
\cA^*_m \gl \veps, \eeqa
where $\veps$ denotes the Majorana supersymmetry transformation
parameter.

\subsection{UnHiggsing the Massive Spin-3/2 Multiplet}
The transformation laws, \pref{E:SusyTransf}, appear to be
singular in the massless limit, $m \to 0$. This is an artifact of
our use of massive fields in this expression--- which amounts to
the use of unitary gauge for the broken supersymmetry and the
broken gauge invariance. The apparent singularity is removed by
unHiggsing this system: that is, by switching to a more convenient
gauge by explicitly identifying the would-be Goldstone modes whose
consumption produced the massive spin-1 and spin-3/2 fields. On
general grounds such a transformation is always possible
\cite{BL,Others}, despite the apparent lack of gauge invariance of
the system.

To perform the unHiggsing we shift the fields $\cA_m$ and $\eta_m$
to expose the eaten scalar, $\phi$, and the eaten goldstino,
$\xi$, as follows:
\eqa
\cA_m \to \cA'_m &=& \cA_m - \, {a \over m} \, \pl_m \phi, \\
\gl \eta_m \to \gl \eta'_m &=& \gl \left[\eta_m + {b\over m}
\,\pl_m \xi + c\, \g_m \xi \right] .\eeqa
The complex constants $a,b,c$ are determined by requiring that the
transformation does not generate off-diagonal kinetic terms in the
lagrangian which mix $\eta_m$ and $\xi$, or $\cA_m$ and $\phi$,
and by requiring the new fields $\phi$ and $\xi$ to have canonical
kinetic terms. These requirements are satisfied by the choice
$a=1$, $b = 2i/\sqrt6$ and $c = i/\sqrt6$, and so we have:
\eqa
\cA'_m &=& \cA_m - \, {1 \over m} \, \pl_m \phi = -\, {1\over m} \cD_m \phi,\\
\eta'_m &=& \eta_m + {2i \over \sqrt{6} m} \, \gf \pl_m \xi + {i
\over \sqrt6} \, \gf \g_m \xi. \eeqa
The second equality in the first line of these equations defines
the scalar covariant derivative $\cD_m \phi = \pl_m \phi - m\cA_m$.

With these choices the lagrangian, eq.~\pref{E:Lkin}, becomes:
\eqa  \label{E:UHLagr}
\cL &=& -\, {i \over 2} \eps^{mnpq} \ebar_m \gf \g_n \pl_p
\eta_q - \, \hf \cA^*_{mn} \cA^{mn} - \, \hf \zbar \dsl \zeta - \,
\hf \xbar \dsl \xi
- \cD_m \phi^* \cD^m \phi \nn\\
&&\qquad  - \, {m \over 2} \, \ebar_m \g^{mn} \eta_n - \, {m\over2} \,
\zbar \zeta - m \, \xbar \xi -im \sqrt{{3\over 2}} \,
(\ebar_m \gf \g^m \xi). \eeqa

We next define the supersymmetry transformations for the new
fields $\phi$ and $\xi$. This is done by partitioning the
supersymmetry transformations of $\cA'_m$ and $\eta'_m$ amongst
the fields $\cA_m$, $\phi$, $\eta_m$ and $\xi$ in such a way as to
ensure that there are no terms proportional to $1/m$. For $\phi$
this gives:
\eq \d \phi = i \sqrt{2} \; \ol{\veps} \gl \left[ {1\over \sqrt3}
(-i \zeta - \sqrt{2} \,\xi) \right], \eeq
which states that the left-handed part of the combination
\eq \chi = {1\over \sqrt3} (-i \gf \zeta - \sqrt{2} \,\xi), \eeq
transforms with $\phi$ with the standard form for $\d\phi$ of a
left-chiral multiplet \cite{Weinberg}. The orthogonal combination
we define as
\eq \lambda = {1 \over \sqrt3} (-\sqrt{2} \,\zeta -i \gf \xi).
\eeq

We similarly choose the transformation rules for $\xi$ to ensure
the absence of $1/m$ terms in $\d\e_m$. We find we must choose:
\eq \gl \d \xi = -\, {i \over 2 \sqrt3} \, \cA^*_{mn} \g^{mn} \gl
\veps + {2i \over \sqrt3} \cD_m \phi \g^m \gr \veps, \eeq
in which case $\lambda$ and $\chi$ transform according to:
\eqa
\gl \d \chi &=& - i \sqrt{2} \;\cD_m \phi \; \g^m \gr \veps, \\
\gl \d \lambda &=& - \, \hf \, \cA^*_{mn} \g^{mn} \gl \veps. \eeqa
Notice that $\cA_m$ does not appear in $\d \chi$ and that $\phi$
drops out of $\d \lambda$. This amounts to the statement that
$\chi$ and $\phi$ form a standard $N=1$ chiral multiplet and
$\lambda$ and $A_m$ form an $N=1$ gauge multiplet.

With these choices we may also write down the transformation law
for $\cA_m$ and $\eta_m$. They are:
\eqa \label{E:SSTrans}
\d \phi &=& i \sqrt{2} \, \ol{\veps} \gl \chi, \nn\\
\gl \d \chi &=& - i \sqrt{2} \; \cD_m \phi \; \g^m \gr \veps, \nn\\
\gl \d \lambda &=& - \, \hf \, \cA^*_{mn} \g^{mn} \gl \veps \\
\d \cA_m &=& \sqrt2 \; \ol\eta_m \gl \veps + \ol{\veps} \g_m \gr \lambda \nn\\
\gl \d \eta_m &=& -\, {1 \over \sqrt2} \cA^-_{mn} \g^n \gr \veps -
\sqrt{2}\; \cD_m \phi^* \, \gl \veps . \nn\eeqa
where $\cA^-_{mn} = \cA_{mn} - \, {i\over 2} \eps_{mnab}
\cA^{ab}$. The lagrangian in these variables becomes:
\eqa \label{E:UHAction}
\cL &=& -\, {i \over 2} \eps^{mnpq} \ebar_m
\gf \g_n \pl_p \eta_q - \, \hf \cA^*_{mn} \cA^{mn} - \, \hf \cbar
\dsl \chi - \, \hf \lbar \dsl \lambda
- \cD_m \phi^* \cD^m \phi \nn\\
&&- \, {m \over 2} \, \ebar_m \g^{mn} \eta_n - \, {m\over 2} \,
\cbar \chi +i \sqrt{2} m \, \lbar \gf \chi + im  \, \ebar_m \gf
\g^m \chi - \, {m \over \sqrt2} \, \ebar_m \g^m \lambda.\nn\\
\eeqa
Notice that both the transformations, eq.~\pref{E:SSTrans}, and
lagrangian density, eq.~\pref{E:UHAction}, now have perfectly
smooth limits as $m\to 0$. Notice also that the fermion mass terms
also have the usual form, with no diagonal gaugino-gaugino pieces,
but only $\chi-\lambda$ mixing plus a $\chi$ mass term.

\subsection{The Nonlinearly Realized Supersymmetry}
We may now identify the second, nonlinearly-realized,
supersymmetry of the action, eq.~\pref{E:UHAction}. Since the
argument closely parallels that of the nonlinearly-realized
internal $U(1)\times U(1)$ gauge invariance of the massive spin-1
particles, we first examine this simpler case.

The nonlinearly-realized $U(1) \times U(1)$ gauge invariance
simply expresses the fact that the two complex fields $\cA_m$ and
$\phi$ only enter the lagrangian through the single combination
$\cA'_m = \cA_m - \pl_m\phi/m$ \cite{BL}. The symmetry therefore
is
\eq \d \phi = m \omega(x), \qquad \d \cA_m = \pl_m \omega , \eeq
where the transformation parameter, $\omega$, is normalized to
ensure $\d\cA_m$ transforms in a conventional way.

The nonlinearly-realized supersymmetry transformation similarly
expresses the fact that the fields $\xi$ and $\eta_m$ only appear
through the single combination $\eta'_m$. Adopting a conventional
normalization for the transformation parameter we therefore have
the second supersymmetry of eq.~\pref{E:UHAction}:
\eq \label{E:NLSSTrans} \d \xi = {i\sqrt{6} \, m\over \kappa}
\; \gf \eta, \qquad \d \eta_m = {2 \over \kappa} \; \pl_m \eta -
\, {m \over \kappa} \g_m \eta . \eeq
This transformation rule for $\xi$ implies $\chi$ and $\lambda$
transform according to
\eq \d \lambda =  {\sqrt{2} \, m \over \kappa} \; \eta, \qquad \d
\chi = - \; {2 i m \over \kappa} \; \gf \eta. \eeq
No other fields transform (to this order in $\kappa$) under this
supersymmetry.

\subsection{Coupling to $N=1$ Supergravity}
Given the globally supersymmetric action just described, we may
immediately write down the coupling to the massless spin-3/2 and
spin-2 fields. This is accomplished using standard methods by
coupling to $N=1$ supergravity. Our interest in what follows is
only in those couplings which arise at lowest order in powers of
$\kappa$.

The fields in the gravity multiplet are the vierbein, ${e_m}^a$,
and massless gravitino, $\psi_m$. The lagrangian density for these
fields is:
\eq \cL_{sg} = - \, {e \over 2 \kappa^2} \; R - \, {i \over 2} \,
\eps^{mnpq} \pbar_m \gf \g_n D_p \psi_q, \eeq
which is invariant with respect to the local supersymmetry
transformation:
\eq \d {e_m}^a = \kappa \, \ol{\veps} \g^a \psi_m, \qquad \d
\psi_m = {2 \over \kappa} \; D_m \veps, \eeq
where $D_m \veps = \pl_m \veps + {1\over 4} {\omega_m}^{ab}
\g_{ab} \, \veps$. The spin connection, ${\omega_m}^{ab}$,
contains gravitino torsion at order $\kappa^2$, but this is
negligible to the order in $\kappa$ to which we work in $\cL$. In
what follows we may therefore treat $D_m$ as if it had no torsion
terms, allowing us to treat the vierbein and Dirac matrices as if
they were covariantly constant. As usual we denote the vierbein
determinant by $e = \det({e_m}^a)$.

The Noether prescription \cite{Noether} can be used to obtain
leading order couplings between the massless $\left(2,\frac32
\right)$ multiplet and the massive $\left(\frac32,1,1,\frac12
\right)$ multiplet \cite{FvN}. The starting point for this
method is the observation that any globally supersymmetric
action transforms under local supersymmetry as:
\eq \d \cL = e (D_m \ol{\veps} ) \, U^m  + \hbox{(total
derivative)}, \eeq
for some Majorana spinor-vector $U^m$. The derivative appearing
here is covariant because we imagine having also already made
$\cL$ generally covariant by replacing all derivatives in the
globally supersymmetric action by covariant derivatives. To
$\cO(\kappa^0)$ this variation is cancelled by the Noether coupling:
$\cL_\kappa = - \, {\kappa \over 2} \, e \, \pbar_m U^m$, due to
the $\cO(1/\kappa)$ variation $\d\psi_m = (2/\kappa)\; D_m \veps$.

Directly varying the lagrangian density, eq.~\pref{E:UHAction}, we
find:
\eqa \d \cL &=& (\hbox{total derivative}) + i \sqrt{2}\, (D_n
\ol\veps \g^m \g^n \gr \chi) \cD_m \phi
 - \cA^{mn}_+ (D_m \ol\veps \g_n \gl \lambda) \nonumber \\
&&  + i \sqrt{2} \eps^{mnpq} (D_m \ol\veps \g_n \gl \eta_p) \,
\cD_q\phi + \, {i\over \sqrt{2}}\, \eps^{mnpq} (D_m \ol\veps \g^r
\g_n \gr \eta_p ) \, \cA^-_{qr} + \cc.\nn \eeqa

We read off from this the supercurrent, $U^m$, for the massive
spin-3/2 multiplet:
\eqa \gr U^m &=& i \sqrt{2}\,( \g^n \g^m \gr \chi) \;\cD_n\phi
- \cA^{mn}_+\, (\g_n \gl \lambda) \nn\\
&& +i \sqrt{2}\, \eps^{mnpq} (\g_n \gl \eta_p) \cD_q \phi +
{i\over \sqrt2} \, \eps^{mnpq} (\g^r \g_n \gr \eta_p) \;
\cA_{qr}^- . \eeqa

Here $\cA_{mn}^\pm = \cA_{mn} \pm {i\over 2} \eps_{mnpq}\cA^{pq}$.
This gives the lagrangian to this order to be $\cL_0 + \cL_\kappa$
where $\cL_0$ contains the kinetic and mass terms:
\eqa \label{E:BA0}
    {\cL_0\over e} &=& - \, {1\over 2 \kappa^2} \; R - \, {i
\over 2e} \, \eps^{mnpq} \pbar_m \gf \g_n D_p \psi_q - \, {i \over
2e} \, \eps^{mnpq}
\ebar_m \gf \g_n D_p \eta_q \nn\\
&&- \, \hf \cA^*_{mn} \cA^{mn} - \, \hf \cbar \dsl \chi - \, \hf
\lbar \dsl \lambda
- \cD_m \phi^* \cD^m \phi \\
&&- \, {m \over 2} \, (\ebar_m \g^{mn} \eta_n) - \, {m\over 2} \,
\cbar \chi +i \sqrt{2} \,m \; (\lbar \gf \chi) + im  \, (\ebar_m
\gf \g^m \chi) - \, {m \over \sqrt2} \, (\ebar_m \g^m \lambda).\nn
\eeqa

The $\cO(\kappa)$ couplings from the Noether prescription are:
\eqa {\cL_\kappa \over e} &=& - \, {i\kappa \over \sqrt2} (\pbar_m
\g^n \g^m \gr \chi) \; \cD_n \phi
+ {\kappa\over 2} \cA^{mn}_+  (\pbar_m \g_n \gl \lambda) \\
&& - {i \kappa \over \sqrt{2} \, e} \eps^{mnpq} \, (\pbar_m \g_n
\gl \eta_p ) \; \cD_q\phi -\, {i \kappa \over 2\sqrt2\, e} \,
\eps^{mnpq} (\pbar_m \g^r \g_n \gr \eta_p) \, \cA^-_{qr} + \cc \nn
\eeqa
The last term in this expression can be simplified to become:
\eqa \label{E:BAk}
    {\cL_\kappa \over e} &=& - \, {i\kappa \over \sqrt2} (\pbar_m
\g^n \g^m \gr \chi) \; \cD_n \phi
+ {\kappa\over 2} \cA^{mn}_+  (\pbar_m \g_n \gl \lambda) \\
&& - {i \kappa \over \sqrt{2} \, e} \eps^{mnpq} \, (\pbar_m \g_n
\gl \eta_p ) \; \cD_q\phi + {\kappa \over \sqrt2} \,(\pbar_m \gr
\eta_n) \, \cA_-^{mn} + \cc \nn \eeqa

The action $\cL = \cL_0 + \cL_\kappa$ is, by construction,
invariant to $\cO(\kappa^0)$ under the unbroken supersymmetry.
Invariance to the same order in $\kappa$ with respect to the
nonlinearly-realized supersymmetry is not automatic, however,
since $\cL_\kappa$ does {\it not} depend on the fields $\phi$,
$\cA_m$, $\xi$ and $\e_m$ only through the invariant combinations
$\cA'_m$ and $\eta'_m$. Furthermore, since the transformation
rules, eq.~\pref{E:NLSSTrans}, contain terms which are
$\cO(1/\kappa)$, the non-invariance of $\cL_\kappa$ already arises
at $\cO(\kappa^0)$ in $\d\cL$. We now show that this non-invariance
may be cancelled by modifying the transformation rules for some of
the fields at $\cO(\kappa^0)$.\footnote{It is with the addition of
these terms that our results differ from those of ref.~\cite{BA}.}

One source of non-invariance is the replacement of ordinary with
covariant derivatives in the globally supersymmetric action and
transformation rules for the second supersymmetry. This ruins the
automatic invariance of the kinetic term for $\eta_m$ to this
order, since the commutator of the two derivatives acting on
$\eta$ is proportional to the Riemann tensor, and so must cancel
the variation of the Einstein term. This requires us to add an
$\cO(\kappa)$ term to $\d {e_m}^a$ involving $\eta_m$, to reproduce
the cancellation of the $\cO(1/\kappa)$ terms which occurs between
the Einstein term and the $\psi_m$ kinetic term. The new
transformation for the vierbein is then symmetric in the two
gravitini:
\eq \d {e_m}^a = \kappa \Bigl( \ol\veps \g^a \psi_m +  \ol\eta
\g^a \eta_m \Bigr). \eeq

In order to cancel the other $\cO(\kappa^0)$ terms in $\d\cL$ we try
the following ansatz, which is the most general consistent with
Lorentz invariance, dimensional analysis and the $U_\ssr(1)$
symmetry:
\eqa \gl \d \psi_m &=& \gl (\d_{\rm old}\psi_m) +
a_1 \, (\gl \eta) \; \cD_m \phi \nonumber\\
&& \qquad + a_2 \, \cA^+_{mn} (\g^n \gr \eta) +
a_3 \, \cA^-_{mn} (\g^n \gr \eta) \nn\\
\d \cA_{m} &=& (\d_{\rm old}\cA_m) + a_4 \, (\pbar_m \gl \eta) \\
\d \phi &=& (\d_{\rm old}\phi) + a_5 \, (\pbar_m \g^m \gr
\eta).\nn \eeqa
The constants $a_1,...,a_5$ are to be chosen to ensure invariance
of $\cL$ up to $\cO(\kappa^0)$. The choice which does so is:
\eq a_1 = \sqrt2, \quad a_2 = 0, \quad a_3 = {1 \over \sqrt2},
\quad a_4 = - \sqrt2, \quad a_5 = 0. \eeq

The combined supersymmetry transformations then are:
\eqa \label{E:BASS}
    \d \phi &=& i \sqrt{2} \, \ol{\veps} \gl \chi, \nn\\
\gl \d \chi &=& - i \sqrt{2} \; \cD_m \phi \; (\g^m \gr \veps) -
2i \, \mu^2 (\gl \eta) \nn\\
\gl \d \lambda &=& - \, \hf \, \cA^*_{mn} (\g^{mn} \gl \veps)
+ \sqrt{2} \, \mu^2 (\gl \eta) \nn\\
\d \cA_m &=& \ol{\veps} \g_m \gr \lambda  +
\sqrt2 \, (\ol\eta_m \gl \veps - \ol\psi_m \gl \eta)  \\
\gl \d \eta_m &=& {2 \over \kappa} \; D_m (\gl \eta)
- \mu^2\g_m (\gr \eta)  \nn\\
&& -\, {1 \over \sqrt2} \, \cA^-_{mn} (\g^n \gr \veps)
- \sqrt{2}\; \cD_m \phi^* \, (\gl \veps) \nn\\
\gl \d \psi_m &=& {2 \over \kappa} D_m (\gl \veps) +\, {1 \over
\sqrt2} \, \cA^-_{mn} (\g^n \gr \eta)
+\sqrt{2}\; \cD_m \phi \, (\gl \eta) \nn\\
\d {e_m}^a &=& \kappa \Bigl( \ol\veps \g^a \psi_m +  \ol\eta \g^a
\eta_m \Bigr) , \nn \eeqa
where $\mu^2 = m/\kappa$.

Eqs.~\pref{E:BA0}, \pref{E:BAk} and \pref{E:BASS} are the main
results of this section, and are generic to the low-energy limit
of {\it any} brane configuration for which supersymmetry is
partially broken from $N=2$ to $N=1$ in four dimensions.

\subsection{Bulk - Brane Split}
Before proceeding, for later purposes it is worth splitting the
fields into two categories which, microscopically, correspond to
those which live on a brane and those which live in the bulk. This
split can be made macroscopically by separating out that part of
the lagrangian which enjoys unbroken $N=2$ supersymmetry, and
assigning this to the bulk.

In order to do so in the present instance we assign to the bulk
the members of the massless $N=2$ supergravity multiplet:
$\{{e^a}_m, \psi_m, \eta_m,B_m \}$. The bulk action consists of
those terms considered previously which depend only on these
fields, and which do not depend on the supersymmetry-breaking
scale $m$ (or $\mu$). Keeping in mind $\cA_m = (A_m + i
B_m)/\sqrt{2}$, we find:
\eqa \label{E:BulkAction}
    {\cL_{B} \over e} &=& - \, {1\over 2 \kappa^2} \; R - \, {i
\over 2e} \, \eps^{mnpq} \pbar_m \gf \g_n D_p \psi_q - \, {i \over2e} \,
\eps^{mnpq} \ebar_m \gf \g_n D_p \eta_q \nn\\
    && \qquad - \, {1 \over 4} \, B_{mn} B^{mn} +  \left[ {i\kappa \over 2}
    \,(\pbar_m \gr \eta_n) \, B_-^{mn} + \cc \right]. \eeqa

Eq.~\pref{E:BulkAction} is invariant under the restriction of the
supersymmetry transformations to these fields, which we call the
bulk part of the transformations:
\eqa \label{E:BulkSusy}
\d_B B_m &=& -i (\ol\eta_m \gf \veps - \ol\psi_m \gf \eta)  \nn\\
\gl \d_B \eta_m &=& {2 \over \kappa} \; D_m (\gl \eta)
-\, {i \over 2} \, B^-_{mn} (\g^n \gr \veps)  \\
\gl \d_B \psi_m &=& {2 \over \kappa} D_m (\gl \veps)
+\, {i \over 2} \, B^-_{mn} (\g^n \gr \eta) \nn\\
\d_B {e_m}^a &=& \kappa \Bigl( \ol\veps \g^a \psi_m +  \ol\eta
\g^a \eta_m \Bigr). \nn\eeqa

All of the remaining terms in the action and supersymmetry
transformations we lump together as the brane contributions, and
so are given to this order in $\kappa$ by
\eqa {\cL_b\over e} &=& - \, {1 \over 4}\, A_{mn} A^{mn} - \, \hf
\cbar \dsl \chi - \, \hf \lbar \dsl \lambda - \cD_m \phi^* \cD^m
\phi - \, {m \over 2} \,
(\ebar_m \g^{mn} \eta_n) \nn\\
&& - \, {m\over 2} \, (\cbar \chi) +i \sqrt{2} \,m \; (\lbar \gf
\chi) + im  \, (\ebar_m \gf \g^m \chi) - \, {m \over \sqrt2} \,
(\ebar_m \g^m \lambda) \\
&&
 - \, {i\kappa \over \sqrt2} (\pbar_m
\g^n \g^m \gr \chi) \; \cD_n \phi
+ {\kappa\over 2} \cA^{mn}_+  (\pbar_m \g_n \gl \lambda) \nn\\
&& - {i \kappa \over \sqrt{2} \, e} \eps^{mnpq} \, (\pbar_m \g_n
\gl \eta_p ) \; \cD_q\phi + {\kappa \over 2} \,(\pbar_m \gr
\eta_n) \, A_-^{mn} + \cc .\nn\eeqa
and
\eqa \label{E:1bSusy}
    \d_b \phi &=& i \sqrt{2} \, \ol{\veps} \gl \chi, \nn\\
\gl \d_b \chi &=& - i \sqrt{2} \; \cD_m \phi \; (\g^m \gr \veps) -2i \,
\mu^2 (\gl \eta) \nn\\
\gl \d_b \lambda &=& - \, \hf \, \cA^*_{mn} (\g^{mn} \gl \veps)
+ \sqrt{2} \, \mu^2 (\gl \eta) \nn\\
\d_b A_m &=& {1 \over \sqrt2} \,  (\ol{\veps} \g_m \lambda)
+ (\ol\eta_m  \veps - \ol\psi_m  \eta)  \\
\d_b B_m &=& {i \over \sqrt2} \, (\ol \veps \g_m \gf \lambda) \nn\\
\gl \d_b \eta_m &=& -\, {1 \over 2} \, A^-_{mn} (\g^n \gr \veps)
- \sqrt{2}\; \cD_m \phi^* \, (\gl \veps) - \mu^2\g_m (\gr \eta)  \nn\\
\gl \d_b \psi_m &=& {1 \over 2} \, A^-_{mn} (\g^n \gr \eta)
+\sqrt{2}\; \cD_m \phi \, (\gl \eta) \nn\\
\d_b {e_m}^a &=& 0. \nn\eeqa

\section{The Single Antibrane}
Before moving on to a multi-brane example, we first pause to ask
how the above construction changes if it is performed for a single
antibrane rather than a single brane.

\subsection{The Brane - Antibrane Map}
Rather than performing the construction again from scratch, we
instead directly write down the result based on the following two
observations.
\begin{itemize}
\item
First, if a brane breaks exactly half of the supersymmetries
of the bulk space of a theory, then typically its antibrane leaves
these supersymmetries unbroken, but breaks the other half of the
supersymmetries.
\item
Second, in explicit microscopic constructions an antibrane
configuration may often be simply obtained from a brane
construction by performing a parity transformation on the
brane world-volume.
\end{itemize}

There are several ways to see the origin of the world-volume
parity transformation. For instance, in four spacetime dimensions
the simplest field theory realization of a brane-like defect might
be the vortex configuration of the abelian Higgs model, which is
characterized by the nonzero magnetic flux which threads the
vortex. The antivortex of this theory has the opposite magnetic
flux, and is obtained simply by rotating the vortex 180$^\circ$
about an axis perpendicular to the vortex. The effect of such a
rotation is simply to reverse the spatial coordinate along the
vortex. Alternatively, a similar argument applies if branes and
antibranes are distinguished by how they couple to antisymmetric
tensors, $B_{m_1,m_2,...}$, which generically are proportional to
$\int B$ over the brane's world volume. The result follows because
this coupling changes sign if we reverse the brane's orientation
--- {\it ie} by a world-volume parity transformation.

These arguments indicate that the antibrane and brane actions
should be obtainable from one another by performing a combination
of a 4D parity transformation and an interchange of the roles of
the two supersymmetries.

The same conclusion may also be drawn directly from the point of
view of the 4D $N=2$ supersymmetry algebra. In this algebra, the
quantity which distinguishes a brane from an antibrane is the
algebra's central charge, $Z$. This enters the 4D $N=2$
commutation relation through
\eq \label{SUSYAlg} \{Q_i, \ol Q_j\} = -2i \g_m P^m \delta_{ij}
           +i\gf \, Z \epsilon_{ij}
\eeq
where $i,j = 1,2$ labels the  two supersymmetries, and
$\epsilon_{ij}$ is the completely antisymmetric tensor. From
eq.~\pref{SUSYAlg} it is clear that the effect of a parity
transformation on the supersymmetry algebra is precisely the same
as the effect of reversing the sign of the central charge.

Notice that the combined operation of parity plus an interchange
of the two supersymmetries leaves the algebra, eq.~\pref{SUSYAlg},
unchanged. It is straightforward to check that it also does not
change the bulk part of the action and supersymmetry transformations,
as these are defined in the previous section. The same is {\it
not} true for the brane parts of the action or supersymmetry
transformations, due to the property that the brane action
realizes one of the supersymmetries linearly and the other
nonlinearly.

A good first start is therefore to define the antibrane action and
supersymmetry transformations by performing a parity
transformation and supersymmetry interchange to the corresponding
quantities for the brane. Actually, these two transformations
alone would imply that the complex scalar, $\phi$, would become
the supersymmetry partner of the right-handed (rather than the
usual left-handed) fermion, so it is also convenient to replace
$\phi \to \phi^*$ and $\chi \to - \chi$ when passing from brane to
antibrane. This ensures that the matter transformations remain in
their standard form. As is straightforward to check, the antibrane
action which results from these replacements is automatically
invariant under the antibrane supersymmetry transformation rules
which result.

The pure bulk part of the action and supersymmetry transformations
are unchanged by the combined action of parity and supersymmetry
interchange, and so are given by
eqs.~\pref{E:BulkAction} and \pref{E:BulkSusy}. The antibrane
action and transformations are directly obtained by performing a
parity transformation and interchanging the fields as discussed
above. Keeping in mind that parity changes the sign of $\gf$ and
$\epsilon^{mnpq}$, and interchanges $\gl \leftrightarrow \gr$ and
$A^\pm_{mn} \leftrightarrow A^\mp_{mn}$, we get in this way:
\eqa {\cL_{\tilde b}\over e} &=& - \, {1 \over 4}\, A_{mn} A^{mn}
- \, \hf \cbar \dsl \chi - \, \hf \lbar \dsl \lambda - \tilde\cD_m
\phi^* \tilde\cD^m \phi - \, {m \over 2} \,
(\pbar_m \g^{mn} \psi_n) \nn\\
&& - \, {m\over 2} \, (\cbar \chi) +i \sqrt{2} \,m \; (\lbar \gf
\chi) + im  \, (\pbar_m \gf \g^m \chi)
- \, {m \over \sqrt2} \, (\pbar_m \g^m \lambda).\nn\\
&& + {i\kappa \over \sqrt2} (\ebar_m \g^n \g^m \gl \chi) \;
\tilde\cD_n \phi^*
+ {\kappa\over 2} \cA^{mn}_-  (\ebar_m \g_n \gr \lambda) \\
&& + {i \kappa \over \sqrt{2} \, e} \eps^{mnpq} \, (\ebar_m \g_n
\gr \psi_p ) \; \tilde\cD_q\phi^* + {\kappa \over 2} \,(\ebar_m
\gl \psi_n) \, A_+^{mn} + \cc .\nn \eeqa
where (because of the change $\phi \to \phi^*$) we have for
antibranes
\eq \tilde\cD_m \phi = \partial_m \phi - m \, \cA^*_m = \partial_m
\phi - \, {m \over \sqrt2} \left( A_m - \, {i } \, B_m \right).
\eeq

The antibrane part of the supersymmetry transformations similarly
becomes
\eqa
\d_b \phi &=& i \sqrt{2} \, \ebar \gl \chi, \nn\\
\gl \d_b \chi &=& - i \sqrt{2} \; \tilde\cD_m \phi \;
(\g^m \gr \eta) - 2i \, \mu^2 (\gl \veps) \nn\\
\gl \d_b \lambda &=& - \, \hf \, \cA_{mn} (\g^{mn} \gl \eta)
+ \sqrt{2} \, \mu^2 (\gl \veps) \nn\\
\d_b A_m &=& {1 \over \sqrt2} \,  (\ebar \g_m \lambda)
- (\ol\eta_m  \veps - \ol\psi_m  \eta)  \\
\d_b B_m &=& - \, {i \over \sqrt2} \, (\ebar \g_m \gf \lambda) \nn\\
\gl \d_b \eta_m &=& +\, {1 \over 2} \, A^-_{mn} (\g^n \gr \veps)
+ \sqrt{2}\; \tilde\cD_m \phi \, (\gl \veps)   \nn\\
\gl \d_b \psi_m &=& -\, {1 \over 2} \, A^-_{mn} (\g^n \gr \eta)
-\sqrt{2}\; \tilde\cD_m \phi^* \, (\gl \eta) - \mu^2 \, (\g_m \gr\veps)\nn \\
\d_b {e_m}^a &=& 0. \nn\eeqa

\section{Two Branes}
Next consider how the above construction changes when there is
more than one brane. We imagine that both of the branes break the
same supersymmetry, and leave the same supersymmetry unbroken.
What we expect to happen in this case would be that the bulk
gravitino which is broken acquires its mass by mixing with a
linear combination of the would-be Goldstone fermions on the two
branes. The examination of this system allows us to see the extent
to which the two branes can remain decoupled --- or sequestered
--- from one another.

The minimal low-energy theory of the two brane configuration may
be constructed by an unHiggsing process, much as was done for the
single brane. The reason for this can be seen by counting the
particle content for the two-brane system. As can be seen from the
one brane system, each brane contributes at least a 4D $N=1$
chiral multiplet, $(\phi,\chi)$, and a gauge multiplet, $(\lambda,
A_m)$. For two branes the low-energy theory contains two of each
of these multiplets. The additional fields have precisely the
content to fill out a single massive $N=1$ spin-1 supermultiplet,
which has the particle content $\left( 1,\frac12,\frac12,0
\right)$. The massive spin-3/2 field and the three massive vector
fields therefore combine with the scalar and spin-1/2 fields into
a massive $N=1$ spin-3/2 multiplet and a massive $N=1$ spin-1
multiplet.

Although the physical spectrum for the two-brane configuration is
a massless $N=1$ spin $\left( 2,\frac32 \right)$ multiplet, a
massive $N=1$ spin $\left( \frac32,1,1,\frac12 \right)$ multiplet,
and a massive $N=1$ spin $\left(1, \frac12, \frac12, 0 \right)$
multiplet, we must still identify which linear combination of
brane and bulk fields combines to form the mass eigenstates. We
may do so by unHiggsing the two massive multiplets as was done
above for the one-brane case.

\subsection{The Massive Spin-1 Multiplet}
To proceed we pause here to record the lagrangian density and
global supersymmetry transformations for the massive $N=1$ spin-1
multiplet. We can then unHiggs it, and rotate to a `brane' basis,
in which the lagrangian and transformation rules look as much as
possible like the sum of terms from two identical branes.

Consider therefore a massive $N=1$ multiplet with spin content
$\left(1,\frac12,\frac12,0\right)$, with the corresponding fields
denoted by $\{A_m,\alpha,\beta,\varphi\}$. If the particle masses
are denoted by $v$, then the free lagrangian density is:
\eqa \cL_v &=& -\, {1\over 4} \, A_{mn} A^{mn} - \, \hf \ol\alpha
\dsl \alpha - \,\hf \, \ol\beta \dsl \beta - \, \hf \partial_m
\varphi
\,\partial^m \varphi \nn\\
&& \qquad - \, {v^2 \over 2} \, A_m A^m - \, {v \over 2} \, (\abar
\alpha + \bbar \beta) - \, {v^2 \over 2} \, \varphi^2. \eeqa

As is shown in Appendix B, this action is invariant with respect
to the global supersymmetry transformations:
\eqa
\d \varphi &=& {1\over \sqrt2} \,\ol\veps \gl(\beta + i \alpha) + c.c. \nn\\
\gl \d \beta &=&{1\over \sqrt2} (\g^m \gr \veps) [\partial_m
\varphi + i v A_m] - {v\over \sqrt2} \,\varphi (\gl \veps) +
{i \over 2\sqrt2}\, A_{mn} (\g^{mn} \gl \veps) \nn\\
\gl \d \alpha &=& -\, {i\over \sqrt2} (\g^m \gr \veps)
[\partial_m \varphi + i v A_m] \\
&&\qquad\qquad-\, {iv\over \sqrt2} \,\varphi (\gl \veps) - \,
{1\over 2\sqrt2} \,
 A_{mn} (\g^{mn} \gl \veps) \nn\\
\d A_m &=& - \partial_m \left[ {i \over v\sqrt2}\, \ol\veps
\gl(\beta + i \alpha) \right] + {i\over \sqrt2} \ol\veps \g_m \gl
(\beta - i \alpha) + c.c. \nn\eeqa

It also shows convenient to define the new (canonically normalized
Majorana) fermions $\chi = (\beta + i\gf\alpha)/\sqrt{2}$ and
$\lambda = (\beta - i\gf \alpha)/\sqrt{2}$, which have more
conventional transformation rules. With these choices we have:
\eqa
\d \varphi &=& \ol\veps \gl\chi + c.c. \nn\\
\gl \d \chi &=&(\g^m \gr \veps)
[\partial_m \varphi + i v A_m] \nn\\
\gl \d \lambda &=& -\, {v} \,\varphi (\gl \veps) +
{i\over 2} \, A_{mn} (\g^{mn} \gl \veps)\\
\d A_m &=& - \partial_m \left[ {i \over v}\, (\ol\veps \gl\chi)
 \right] + {i} (\ol\veps \g_m \gl \lambda)
+ c.c.  \nn \eeqa
This result may be unHiggsed by shifting $A_m^{\rm old} = A_m^{\rm
new} - \, {1 \over v} \, \partial_m \sigma$, and requiring all
$1/v$ terms to vanish as a result. We find the unHiggsed
lagrangian is
\eqa \cL_v &=& -\, {1\over 4} \, A_{mn} A^{mn} - \, \hf \lbar \dsl
\lambda - \,\hf \, \cbar \dsl \chi - \, \hf \partial_m \varphi
\,\partial^m \varphi  \nn\\
&& \qquad - \, \hf \, \cD_m \sigma \, \cD^m \sigma
 - \, {v} \,
(\cbar \lambda) - \, {v^2 \over 2} \, \varphi^2, \eeqa
where $\cD_m \sigma = \partial_m \sigma - v A_m$.

Writing $\phi = (\sigma + i \varphi)/\sqrt2$ we have the
supersymmetry transformations
\eqa \label{E:spin1Susy}
    \d \phi &=& i \sqrt{2} \, (\ol\veps \gl\chi) \nn\\
\gl \d \chi &=& -i\sqrt{2} \, (\g^m \gr \veps)
\left[\partial_m \phi - \, { v\over \sqrt{2}} \,  A_m \right] \nn\\
\gl \d \lambda &=&  {iv\over \sqrt{2}} \, (\phi- \phi^*) (\gl
\veps) + {i\over 2}\, \, A_{mn}
(\g^{mn} \gl \veps) \\
\d A_m &=&  i (\ol\veps \g_m \gl \lambda) + c.c. = i \ol\veps \g_m
\gf \lambda \nn \eeqa
These transformation rules suggest the definition $\cD_m\phi =
\partial_m \phi - {v\over\sqrt2}\, A_m$.

\subsection{The Two-Brane System}
We now couple the spin-3/2 and spin-1 supermultiplets to the
massless spin-2 multiplet of unbroken $N=1$ supergravity. We
imagine that the branes are identical, and so contribute
equally to the breaking of the broken supersymmetry.

To this end, we take the bulk matter content to be
$\{{e^a}_m,\psi_m,\eta_m,B_m\}$, and supplement this with two
copies of the brane matter: $\{A_m,\lambda, \chi,\phi\}$ and
$\{A'_m, \lambda', \chi', \phi' \}$. In general these fields
combine into a massive spin-3/2 multiplet plus a massive spin-1
multiplet, whose masses need not be equal. We consider the case
where both multiplets have the same mass, which we denote by $m$.

Under these assumptions the brane lagrangian is given by the
following expression when written in terms of mass eigenstates:
$\cL = \cL_B + \cL_{2b}$, where $\cL_B$ is the bulk lagrangian
given earlier, eq.~\pref{E:BulkAction}. The brane part of the
action is the sum of the massive spin-3/2 and spin-1 lagrangians:
$\cL_{2b} = \cL_{3/2} + \cL_{1}$, where
\eqa {\cL_{3/2}\over e} &=& - \, {1 \over 4}\, A_{mn} A^{mn} - \,
\hf \cbar \dsl \chi - \, \hf \lbar \dsl \lambda - \cD_m \phi^*
\cD^m \phi - \, {m \over 2} \,
(\ebar_m \g^{mn} \eta_n) \nn\\
&& - \, {m\over 2} \, (\cbar \chi) +i \sqrt{2} \,m \; (\lbar \gf
\chi) + im  \, (\ebar_m \gf \g^m \chi)
- \, {m \over \sqrt2} \, (\ebar_m \g^m \lambda), \nn\\
\eeqa
and
\eqa \label{E:Spin1Action}
    \cL_1 &=& -\, {1\over 4} \, A'_{mn} A^{'mn} - \, \hf \lbar'
\dsl \lambda' - \,\hf \, \cbar' \dsl \chi'
- \, \cD_m {\phi'}^* \, \cD^m \phi' \nn\\
&& \qquad  - \, {m } \, ( \cbar' \lambda') + \, {m^2\over 4} \,
(\phi' - {\phi'}^*)^2, \eeqa
where
\eq \cD_m \phi = \partial_m \phi - \, {m \over \sqrt{2}} \, (A_m +
i B_m), \qquad \cD_m \phi' = \partial_m \phi' - \, {m \over
\sqrt{2}} \, A'_m . \eeq

Both $\cL_{3/2}$ and $\cL_1$ have the supersymmetries we have
worked out, with the fields in $\cL_{3/2}$ transforming as in
eqs.~\pref{E:1bSusy} and those in $\cL_1$ transforming as in
eqs.~\pref{E:spin1Susy}. We wish to now rotate the matter
multiplets to a brane basis, for which the massless limit is
nonsingular and the two brane contributions look similar to one
another. Since we assume both branes contribute equally to the
breaking of the supersymmetry, we can assume a symmetry under
their interchange, and so we can assume that the required rotation
between the brane states and the mass eigenstates is through 45
degrees. We therefore take:
\eqa \label{E:2bRot}
    \pmatrix{A_m \cr A'_m \cr} = R \; \pmatrix{A^1_m \cr A^2_m},
&\quad& \pmatrix{\lambda \cr \lambda' \cr} = R \;
\pmatrix{\lambda_1
\cr \lambda_2}, \nn\\
\pmatrix{\chi \cr \chi' \cr} = R \; \pmatrix{\chi_1 \cr \chi_2},
&\quad& \pmatrix{\phi \cr \phi' \cr} = R \; \pmatrix{\phi_1 \cr
\phi_2}, \eeqa
with
\eq R = \pmatrix{c_\theta & s_\theta \cr -s_\theta & c_\theta} =
{1 \over \sqrt{2}} \; \pmatrix{1 & 1 \cr -1 & 1 \cr} \eeq
where $c_\theta = \cos\theta$ and $s_\theta = \sin\theta$.

Notice if we define:
\eq \cD_m \phi_1 = \partial_m \phi_1 - \, {m \over \sqrt2}
\left(A^1_m + {i \over \sqrt 2} \, B_m \right), \qquad \cD_m
\phi_2 = \partial_m \phi_2 - \, {m \over \sqrt2} \left(A^2_m + {i
\over \sqrt 2} \, B_m \right), \eeq
then we also have:
\eq \pmatrix{\cD_m \phi \cr \cD_m \phi' \cr} = R \pmatrix{\cD_m
\phi_1 \cr \cD_m \phi_2 \cr}. \eeq

The brane action is obtained by inserting this rotation into the
above lagrangian, and the supersymmetries are similarly found by
using this rotation in the known supersymmetry transformations. In
this way we find $\cL_{2b} = \cL_{b1} + \cL_{b2} + \cL_{b1b2}$
with:
\eqa \cL_{b1} &=& -\, {1\over 4} \, A^1_{mn} A_1^{mn} - \, \hf
\lbar_1 \dsl \lambda_1 - \,\hf \, \cbar_1 \dsl \chi_1
- \, \cD_m \phi_1^* \, \cD^m \phi_1 \nn\\
&& \qquad  + {im \over \sqrt2} \, (\ebar_m \gf \g^m \chi_1) - \,
{m \over 2} \, (\ebar_m \g^m \lambda_1) -\, {m\over 4}
\, (\ebar_m \g^{mn} \eta_n) \\
&& \qquad + {m^2 \over 8} \, (\phi_1 - \phi_1^*)^2 -\, {m \over
4}\, \Bigl[ (\cbar_1 \chi_1) + 2 \, \cbar_1(1 - i \sqrt{2}\, \gf)
\lambda_1 \Bigr],\nn \eeqa
and $\cL_{b2}$ obtained from this by replacing $1 \to 2$
everywhere. The direct brane-brane mixing terms are:
\eqa \label{E:2bcontact}
    \cL_{b1b2} &=& - \, {m^2 \over 4} \, (\phi_1 - \phi_1^*)
(\phi_2 - \phi_2^*) \\
&& \qquad - \, {m \over 2} \, \Bigl[(\cbar_1 \chi_2) - \cbar_1(1 +
i \sqrt{2} \, \gf) \lambda_2 - \cbar_2 (1 + i \sqrt{2} \, \gf)
\lambda_1 \Bigr].\nn \eeqa

The supersymmetry transformation laws are similarly obtained by
substitution of the field rotation into those rules which were
previously derived. One finds the brane-field-dependent part of
the bulk-field transformation laws to be
\eqa \d_b B_m &=& {i \over 2} \, \ol \veps \g_m \gf
(\lambda_1 + \lambda_2) \nn\\
\gl \d_b \eta_m &=& -\, {1 \over 2\sqrt2} \,
(A^{1-}_{mn}+A^{2-}_{mn}) (\g^n \gr \veps)\\
&& \qquad- (\cD_m \phi_1^* + \cD_m \phi_2^*)
\, (\gl \veps) - \, {m \over \kappa}\, \g_m (\gr \eta) \nn \\
\gl \d_b \psi_m &=& {1 \over 2\sqrt2} \, (A^{1-}_{mn}+A^{2-}_{mn})
(\g^n \gr \eta)  + (\cD_m \phi_1 + \cD_m\phi_2)
\, (\gl \eta) \nn\\
\d_b {e_m}^a &=& 0. \nn\eeqa

The transformations of the brane fields become:
\eqa
\d \phi_1 &=& i \sqrt{2} \, \ol{\veps} \gl \chi_1, \nn\\
\gl \d \chi_1 &=& - i \sqrt{2} \; \cD_m \phi_1 \; (\g^m \gr \veps)
-i\sqrt{2}  \, {m \over \kappa} (\gl \eta) \nn\\
\gl \d \lambda_1 &=& -\, {1 \over 4\sqrt2} \left[(1 -i\sqrt2)\,
A^1_{mn} + (1+i\sqrt2) \, A^2_{mn} -i \sqrt{2}\,B_{mn} \right]
(\g^{mn} \gl \veps)
\nn\\
&& \qquad +{im\over 2\sqrt2} (\phi_1 - \phi_2 - \phi_1^* +
\phi_2^*)
(\gl \veps) + {m\over \kappa} \, (\gl \eta) \\
\d A^1_m &=& {1 \over 2\sqrt2} \, \ol\veps \g_m (1 + i \sqrt{2} \,
\gf) \lambda_1 + {1 \over 2 \sqrt2} \, \ol\veps \g_m (1 - i
\sqrt{2} \, \gf)
\lambda_2 \nn\\
&& \qquad\qquad\qquad + {1\over \sqrt2} \, (\ebar_m \veps -
\pbar_m \eta). \nn\eeqa
with the transformation rule for brane-2 fields obtained by making
the substitutions $1 \leftrightarrow 2$ in these expressions.

\subsection{Noether Coupling to Two Branes}
The couplings to supergravity which arise from the Noether
prescription are again easily identified. Starting with the
massive spin-1 multiplet, eq.~\pref{E:Spin1Action}, inserting
covariant derivatives throughout and following the derivatives of
the supersymmetry parameter in the variation of the action, one
finds:
\eqa \label{E:Spin1kappa}
    {\cL_{1\kappa} \over e} &=& {i \kappa \, m\over 2 \sqrt2} \;
    \pbar_m \g^m \gl \lambda' \, (\phi' - \phi'{}^*) + {i \kappa
    \over 2} \, \pbar_m \g_n \gl \lambda' \, A_+'{}^{mn} \nn\\
    && \qquad\qquad + {i \kappa\over \sqrt2} \, \pbar_m \g^n \g^m \gl \chi' \,
    \cD_n \phi'{}^* + \cc \, .
\eeqa
This, when combined with eq.~\pref{E:BAk}, gives the complete
$\cO(\kappa)$ Noether couplings for the two brane system.

Substitution of the rotations of eq.~\pref{E:2bRot} into these
expressions gives them in terms of the brane fields $\phi_1$,
$\phi_2$ {\it etc.}. Once this is done it is clear that terms
arise at $\cO(\kappa)$ which contribute to the direct couplings
between fields arising on each brane.

\section{Conclusions}
We have studied the form of the low-energy action for brane
configurations which partially break 4D $N=2$ supersymmetry down
to 4D $N=1$ supersymmetry. Our focus has been on those fields
which are necessarily present in this kind of symmetry breaking
pattern, because they are required in order to fill out the
required $N=1$ supersymmetry multiplets which contain the massive
spin-3/2 and spin-1 fields which arise. We have seen how the
linear realization of one supersymmetry and the nonlinear
realization of the other strongly constrain the form the resulting
low-energy action can take.

In particular, our analysis allows us to address the sequestering
conjecture of ref.~\cite{RS0}, in which a strong decoupling of the
degrees of freedom on two supersymmetric branes is proposed. We
can test a part of the sequestered form for the K\"ahler
potential, eq.~\pref{E:sequester}, since this implies that the
terms which directly mix the chiral multiplets of the two branes
arise for the first time at $\cO(\kappa^2)$ in the effective
lagrangian.

We find that this kind of sequestered decoupling is not satisfied
by the degrees of freedom we follow in the low-energy theory,
since direct brane-brane contact couplings arise already at
$\cO(\kappa)$, as may be seen from eq.~\pref{E:2bcontact} if we
use the natural relation $m = \kappa \mu^2$ to see that $m$ itself
is $\cO(\kappa)$ relative to the size, $\mu^2$, of the
supersymmetry-breaking order parameter. A similar conclusion
follows from an inspection of the Noether couplings for two
branes.

These findings complement and reinforce those of ref.~\cite{ADGT},
who examined the couplings between branes in specific types of
string theory compactifications for which direct brane-brane
couplings are explicitly calculable and attributable to exchanges
of bulk supergravity fields, or to the warping of the internal
compact dimensions. The present analysis shows that these contact
terms do not depend on the details of these compactifications, but
follow quite robustly from the assumed 4D pattern of partial
supersymmetry breaking.

The phenomenological appeal of low-energy sequestering for
circumventing the flavor problems of low-energy supersymmetry
breaking makes the continued search for models with sequestered
sectors well worthwhile. Our results show that any such system
cannot be obtained simply by coupling additional fields which do
not participate in the diagonalization of the gravitino masses.
Instead a more radical change is required, and it may be that what
is required is a supersymmetry-breaking pattern within which a
larger extended supersymmetry is partially broken, such as perhaps
$N=4$ breaking to $N=2$. Some evidence that this might be the case
can be found from the expression of these extended supersymmetries
in terms of $N=1$ superfields \cite{JP}, which are often
suggestive of sequestering.

Our techniques can clearly be extended to the complete breaking of
$N=2$ supersymmetry, such as is obtained by combining the
supersymmetry breaking of one brane with that of an antibrane. In
such a system we naturally expect that the full $N=2$
supersymmetry will be nonlinearly realized, and the challenge is
to find what additional constraints distinguish the resulting
low-energy theory from a generic non-supersymmetric model. We see
the present work as providing the first step towards this more
ambitious project of studying the low-energy implications of
supersymmetry breaking in the brane world. This is of particular
relevance given the fact that there are explicit constructions of
D-brane models for which either supersymmetry is broken by
brane/antibrane systems, or in intersecting brane models where
different intersecting branes break different supersymmetries,
whereas the bulk preserves higher supersymmetries. This kind of
`non-local' realization of supersymmetry breaking by many branes
in a supersymmetric bulk was proposed in \cite{BMQ} as a way to
ameliorate the cosmological constant problem, and has been
recently constructed in intersecting D-brane models \cite{CIM},
where it was called quasi-supersymmetry (Q-SUSY). It is clearly of
great interest to study the low-energy limits of these models and
uncover whether the brane constructions lead to new low-energy
mechanisms for keeping parameters naturally small. We hope to
report on some of these issues in future publications.

\acknowledgments We thank L.~Ib\'a\~nez for interesting
conversations. C.B.'s research is partially funded by grants from
N.S.E.R.C. of Canada and F.C.A.R. of Qu\'ebec. C.B. and F.Q.
respectively thank D.A.M.T.P. of Cambridge University and the CERN
Theory Division for their hospitality during various parts of this
work. M.K. wants to thank R.~Altendorfer for useful e-mail
correspondence. The work of M.K. is currently funded by the
Deutsche Forschungsgemeinschaft, but the major part of this work
was performed while M.K. was an SPG fellow at the University of
Cambridge supported by PPARC. F.Q. is partially supported by
PPARC.

\begin{appendix}
\section{Conventions}
Since supersymmetric calculations tend to be finicky, and since
conventions vary across the supersymmetry literature, we spell out
our conventions in detail in this appendix.

\subsection{Majorana spinor formalism}
We use Majorana spinors throughout, with 4-component gamma
matrices. The notation closely follows that of Weinberg's textbook
\cite{Weinberg} --- with the exception of the sign we use for the
gravitino mass term --- and is similar to that used in West's book
\cite{West}. In particular we use Weinberg's curvature
conventions, which differ by an overall minus sign from the
curvature conventions of West and of Misner, Thorne and Wheeler
\cite{MTW}.

\bigskip{\bf Basic Definitions}
$$\hbox{metric:} \qquad \eta^{mn} = \diag(-+++)$$
$$\hbox{epsilon tensor:} \qquad \epsilon^{0123} = -\epsilon_{0123} = 1$$
$$ \epsilon_{mnpq} \epsilon^{mnab} = -2 \left( \d^a_p \d^b_q - \d^a_q \d^b_p\right) \qquad
\epsilon_{mnpq} \epsilon^{mabc} = - \left( \d^a_n \d^b_p \d^c_q
\pm 5 \;\hbox{perms}\right) $$
$$\hbox{Dual Tensors:} \qquad \tilde{t}_{mn} = {i \over 2} \epsilon_{mnab} t^{ab} $$
$$\hbox{(Anti) Selfdual Tensors:} \qquad t_\pm^{mn} = t^{mn} \pm {i \over 2}
\epsilon^{mnab} t_{ab} = t^{mn} \pm \tilde{t}^{mn}\qquad \qquad
\tilde{t}_\pm^{mn} = \pm t_\pm^{mn}$$
$$\hbox{Dirac matrices:} \qquad \{\g^m, \g^n\} = 2\eta^{mn}$$
$$\hbox{Commutator:} \qquad \g^{mn} = {1\over 2} [\g^m, \g^n]$$
$$\hbox{Projectors:} \qquad \gl = {1\over 2} (1 + \gf), \qquad \gr = {1\over 2}(1 - \gf)$$

\bigskip{\bf Flat-Space Dirac Identities}
$$ \g_m\g_n = \e_{mn} + \g_{mn} $$
$$\gf \g_{mn}  = -\, {i \over 2} \epsilon_{mnpq} \g^{pq}$$
$$ \epsilon^{mnpq} \gf \g_{na} = -i \left( \d^m_a \g^{pq} - \d^p_a \g^{mq}
+ \d^q_a \g^{mp} \right) $$
$$ \epsilon^{mnab} \gf \g_{ab} = 2i \; \g^{mn} $$
$$\g_m \g_{np} = (\e_{mn} \g_p - \e_{mp} \g_n) + i \epsilon_{mnpq} \gf \g^q $$
$$\g_{np} \g_m = -(\e_{mn} \g_p - \e_{mp} \g_n) + i \epsilon_{mnpq} \gf \g^q $$
$$ \g_m\g_n\g_p = \g_m \e_{np} + \g_p \e_{mn} - \g_n \e_{mp} + i \epsilon_{mnpq} \gf \g^q $$
$$ \g_{mn} \g_{pq} = \e_{mq} \e_{np} - \e_{mp} \e_{nq} + i \gf \epsilon_{mnpq}
- \e_{mp} \g_{nq} + \e_{np} \g_{mq} - \e_{nq} \g_{mp} + \e_{mq}
\g_{np} $$
$$ [\g_{mn}, \g_{pq}] = 2 \left( - \e_{mp} \g_{nq} + \e_{np} \g_{mq} - \e_{nq} \g_{mp}
+ \e_{mq} \g_{np} \right) $$
$$ \{\g_{mn}, \g_{pq}\} = 2 \left( \e_{mq} \e_{np} - \e_{mp} \e_{nq} +
i \gf \epsilon_{mnpq}\right) $$

\bigskip{\bf Lorentz Transformations}
$$\hbox{Lorentz Generators:} \qquad \hbox{4-vectors:} \qquad \delta V_m = {i\over 2}
\omega^{ab} {(J_{ab})_m}^n V_n = {\omega_m}^n V_n$$
$$\hbox{This implies for 4 vectors:} \qquad {(J_{ab})_m}^n = -i \left( \eta_{am}
\delta_b^n - \e_{bm} \d^n_a \right)$$
$$\hbox{This fixes the Lorentz Commutators:} \qquad  [J_{ab},
J_{cd}] = -i \e_{bc} J_{md} \pm \hbox{permutations}$$
$$\hbox{Spinor Lorentz generators:} \qquad  \delta \psi = {i\over 2}
\omega^{ab} J_{ab} \psi, \qquad J_{ab} = -\, {i\over 2} \g_{ab}$$
$$\hbox{Spinor Covariant Derivative:} \qquad  D_m \psi = \partial_m\psi
+ {i\over 2} \omega_m^{ab} J_{ab} \psi = \partial_m \psi + \,
{1\over 4} {\omega_m}^{ab} \g_{ab} \psi$$
$$\hbox{Curvature Commutator - spinor:} \qquad  [D_m,D_n] \psi = -\, {i\over 2}
{R_{mn}}^{ab} J_{ab} \psi = -\, {1\over 4} {R_{mn}}^{ab} \g_{ab}
\psi$$
$$\hbox{Curvature Commutator - vector:} \qquad  [D_m,D_n] V_p = -\, {i\over 2}
{R_{mn}}^{ab} {(J_{ab})_p}^q V_q = {R_{mnp}}^{q} V_q$$

\bigskip{\bf Majorana Spinors} If $\chi$ and $\psi$ are any Majorana spinors, then:
$$\hbox{Reality:} \qquad (\ol{\chi} M \psi)^* = \pm \ol\chi M \psi ; \qquad
\hbox{with + for $M = S,V,T$ and $-$ for $M=P,A$}  $$
$$\hbox{Symmetry:} \qquad \ol{\chi} M \psi = \pm \ol\psi M \chi ; \qquad
\hbox{with + for $M=S,P,A$ and $-$ for $M=V,T$}  $$

\subsection{Weyl spinor formalism}
We would like to show how the results of this article can be translated
into the Weyl spinor formalism of Wess and Bagger \cite{WB}.
The Weyl spinor components of a Dirac 4-spinor $\Psi$ are defined through
\eq  \label{Weyl_Dirac}
\psi\equiv\ol{\gl\Psi}=\ol{\Psi}\gr,\qquad
\chi\equiv\gr\Psi.
\eeq
If $\Psi$ is restricted to satisfy a Majorana condition, then $\chi=\psi$.
In this case, $\Psi$ can be written as
\eq  \label{two_four}
\Psi_a=\left(\ba{c}\bar\psi^{\dot\alpha}\\ \psi_\alpha\ea\right)
\qquad\Rightarrow\qquad
\ol\Psi^a=\left(\bar\psi_{\dot\alpha},\,\psi^\alpha\right),
\eeq
where $a=(\dot\alpha,\alpha)$ is a 4-spinor index. Similarly, the
$\gamma$-matrices can be split into blocks of $\sigma$-matrices:
\eq  \label{gamma_sigma}
\left(\gamma^m\right)_a^{\ b}=i\left(\ba{cc}
0 &(\bar\sigma^m)^{\dot\alpha\beta}\\
(\sigma^m)_{\alpha\dot\beta} &0  \ea\right),\qquad
\left(\gamma^{mn}\right)_a^{\ b}=-2\left(\ba{cc}
(\bar\sigma^{mn})^{\dot\alpha}_{\ \dot\beta} &0\\
0 &(\sigma^{mn})_\alpha^{\ \beta}  \ea\right).
\eeq

To make contact with the notation of \cite{BA}, we define two Weyl spinor
gravitini $\psi^{(1)}$, $\psi^{(2)}$ through
\eq  \label{Weyl_gravitino}
\eta_m=i\gamma_5\left(\ba{c}\bar\psi^{(2)}_m\\ \psi^{(2)}_m\ea\right),\qquad
\psi_m=\left(\ba{c}\bar\psi^{(1)}_m\\ \psi^{(1)}_m\ea\right).
\eeq
The chiral rotation of the second gravitino was necessary because we have
chosen an unconventional sign for the gravitino mass term in our lagrangians.
Similarly, we define the Weyl spinor supersymmetry transformation parameters
$\eta^{(1)}$, $\eta^{(2)}$ through
\eq  \label{Weyl_susyparam}
\veps=\left(\ba{c}\bar\eta^{(1)}\\ \eta^{(1)}\ea\right),\qquad
\eta=i\gamma_5\left(\ba{c}\bar\eta^{(2)}\\ \eta^{(2)}\ea\right).
\eeq
The Weyl spinor Goldstino $\nu$ is defined by
\eq  \label{Weyl_goldstino}
\nu=-\gr\xi.
\eeq
In this section, we denote by $\zeta$, $\chi$, $\lambda$ the Weyl spinor
components of the Majorana spinors $\zeta$, $\chi$, $\lambda$ in the
main text, {\it ie}, we replace
\eq  \label{Weyl_zeta}
\gr\zeta\to\zeta,\qquad
\gr\chi\to\chi,\qquad
\gr\lambda\to\lambda.
\eeq
Finally, we replace
\eq  \label{Aphi_replace}
\cA_m\to{i\over\sqrt2}\,\bar\cA_m,\qquad
\phi\to i\bar\phi.
\eeq

In this notation the unHiggsed massive gravitino lagrangian
\pref{E:UHLagr} reads
\eqa  \label{unHiggsed}
\cL &= &\eps^{mnpq}\bar\psi^{(2)}_m\bar\sigma_n\partial_p\psi^{(2)}_q
        -i\,\bar\nu\bar\sigma^m\partial_m\nu
        -i\,\bar\zeta\bar\sigma^m\partial_m\zeta
        -\cD^m\phi\bar\cD_m\bar\phi
        -\qt\cF^{mn}\bar\cF_{mn}       \nonumber\\
    &  &-m\left(\psi^{(2)}_m\sigma^{mn}\psi^{(2)}_n
            +{3i\over\sqrt6}\psi^{(2)}_m\sigma^m\bar\nu
            +\nu\nu+\hf\zeta\zeta\ +\ h.c.\right),
\eeqa
where $\cF_{mn}=\partial_{[m}\cA_{n]}$ and
$\cD_m\phi=\partial_m\phi-{m\over\sqrt2}\cA_m$.

The corresponding transformation laws are
\eqa \label{st_new}
\delta_\eta \psi_m &= &-{i\over2}\cF_{+mn}\sigma^n\bar\eta
                            +\sqrt2\,\bar\cD_m\bar\phi\,\eta \nonumber\\
\delta_\eta \nu    &= &-{1\over\sqrt6}\left(\bar\cF_{mn}\sigma^{mn}\eta
                            -2\sqrt2\,i\,\cD_n\phi\,\sigma^n\bar\eta\right)
                                                           \nonumber\\
\delta_\eta \cA_m  &= &2\,\psi_m\eta
                       -{2\,i\over\sqrt6}\left(-i\sqrt2\,\bar\zeta
                              -\bar\nu\right)\bar\sigma_m\eta
                                                            \nonumber\\
\delta_\eta \phi   &=&\sqrt{2\over3}\left(i\zeta+\sqrt2\,\nu\right)\eta
                                                            \nonumber\\
\delta_\eta \zeta  &= &-{i\over\sqrt3}\,\bar\cF_{mn}\sigma^{mn}\eta
                       +\sqrt{2\over3}\,\cD_m\phi\,\sigma^m\bar\eta
\eeqa

The Noether coupling of the massive gravitino lagrangian to supergravity
yields
\eqa  \label{L_nlsugra}
e^{-1}\cL &= &-{1\over2\kappa^2}R+\eps^{mnpq}\bar\psi^{(i)}_m
                     \bar\sigma_nD_p\psi^{(i)}_q
              -i\,\bar\chi\bar\sigma^mD_m\chi
              -i\,\bar\lambda\bar\sigma^mD_m\lambda \nonumber\\
          &  &-\cD^m\phi\bar\cD_m\bar\phi-\qt\cF^{mn}\bar\cF_{mn}
                             \nonumber\\
          &  &-m\left({1\over\sqrt2}\psi^{(2)}_m\sigma^m\bar\lambda
                  +i\,\psi^{(2)}_m\sigma^m\bar\chi
                  +i\sqrt2\,\lambda\chi+\hf\chi\chi
                  +\psi^{(2)}_m\sigma^{mn}\psi^{(2)}_n\right)
                             \nonumber\\
          &  &-{\kappa\over\sqrt2}\left(\chi\sigma^m\bar\sigma^n\psi^{(1)}_m
                  \bar\cD_n\bar\phi+\hf\bar\lambda\bar\sigma_m\psi^{(1)}_n
                  \bar\cF_-^{mn}+\eps^{mnpq}\bar\psi^{(2)}_m\bar\sigma_n
                  \psi^{(1)}_p\bar\cD_q\bar\phi\right) \nonumber\\
          &  &+{\kappa\over4}\eps^{ij}\psi^{(i)}_m\psi^{(j)}_n\bar\cF_+^{mn}
              \ +\ h.c.,
\eeqa
which is the Weyl spinor expression of the lagrangian \pref{E:BA0},
\pref{E:BAk}. Here, we defined
\eq  \label{chi_lambda}
\chi\equiv{1\over\sqrt3}(i\zeta+\sqrt2\nu),\qquad
\lambda\equiv{1\over\sqrt3}(-\sqrt2\zeta-i\nu).
\eeq

The corresponding transformation laws \pref{E:BASS} are
\eqa \label{st_nlsugra}
\delta_\eta e^a_{\ m}  &= &i\,\kappa\left(\eta^{(i)}\sigma^a
                             \bar\psi^{(i)}_m+\bar\eta^{(i)}
                             \bar\sigma^a\psi^{(i)}_m\right),\nonumber\\
\delta_\eta \psi^{(i)}_m &=&{2\over\kappa}D_m\eta^{(i)}
                  +{i\over2}\eps^{ij}\cF_{+mn}\sigma^n\bar\eta^{(j)}
                  +\sqrt2\,\left(\bar\cD_m\bar\phi\eta^{(1)}\delta^{i2}
                              -\cD_m\phi\eta^{(2)}\delta^{i1}\right)
                                                            \nonumber\\
                   &  &+i\,\mu^2\sigma_m\bar\eta^{(2)}\delta^{i2},
                                  \nonumber\\
\delta_\eta \cA_m  &= &-2\eps^{ij}\psi^{(i)}_m\eta^{(j)}
                           +\sqrt2\,\bar\lambda\bar\sigma_m\eta^{(1)},
                                  \nonumber\\
\delta_\eta \lambda &= &{i\over\sqrt2}\bar\cF_{mn}\sigma^{mn}
                          \eta^{(1)}-i\sqrt2\,\mu^2\eta^{(2)},
                                  \nonumber\\
\delta_\eta \chi    &= &i\sqrt2\,\sigma^m\cD_m\phi\bar\eta^{(1)}
                       +2\mu^2\eta^{(2)},    \nonumber\\
\delta_\eta \phi    &= &\sqrt2\,\chi\eta^{(1)},
\eeqa
where one has the relation $m=\kappa\mu^2$.

\section{Derivation of Supersymmetry Transformations}
In this Appendix we derive the global supersymmetry transformation
rules for the massive spin-3/2 and spin-1 multiplets.

\subsection{The Massive Spin-3/2 Multiplet}
For the spin-3/2 multiplet we start with the following ansatz for
the multiplet's supersymmetry transformation rules, which consists
of the most general form permitted by Lorentz invariance and
dimensional analysis, provided that any $1/m$ terms are pure
gauge, so that they do not introduce $1/m$ terms into $\d
\cL_{\rm kin}$. We also require the ansatz to be $U(1)$ invariant
where $\cA_m \to e^{i\alpha} \, \cA_m$ and $\gl \veps \to
e^{i\alpha} \;\gl \veps$. This leads to the following form:
\eqa \d \cA_m &=& c_1 (\ol\eta_m \gl \veps) + c_2 (\zbar \g_m \gl
\veps) + {c_3\over m}
\pl_m \Bigl( \zbar \gl \veps \Bigr), \nn\\
\gl \d \zeta &=& c_4 \cA^*_{mn} \; \g^{mn} \gl \veps + c_5 m \cA_m \, \g^m \gr \veps, \\
\gl \d \eta_m &=& {1 \over m} \pl_m \left( c_6 \cA_{ab}^* \g^{ab}
\gl \veps
+ m c_7 \cA_a \g^a \gr \veps \right) + c_8 \cA_{mn} \g^n \gr \veps \nn\\
&& + c_9 \eps_{mnab} \cA^{ab} \g^n \gr \veps + c_{10} m \cA^{n*}
\g_{mn} \gl \veps + c_{11} m \cA^*_m \gl \veps. \nn \eeqa

Demanding $\d \cL = 0$, order by order in $m$ gives the
conditions:
\eq c_2 = 2 c_4, \quad c_3 = - 2 c_4, \quad c_5 = -2 c_4^* \eeq
coming from those terms in $\d \cL$ which involve $\zeta$, and the
rest give:
\eq c_6 = {c_1\over 6}, \quad c_7 = {2 c_1^*\over 3}, \quad c_8 =
- \, {2 c_1^*\over 3}, \quad c_9 = {i c_1^* \over 6}, \quad c_{10}
= - \, {c_1\over 3}, \quad c_{11} = {2 c_1\over 3}. \eeq
The constants $c_1$ and $c_4$ are not determined. They could be
fixed by closing the algebra and asking the translation generated
to be normalized in the standard way. Instead we choose them by
comparing with ref.~\cite{BA}, with the idea of making contact
with a conventional form for the transformation laws for a later
choice of variables. This leads to the choices (keeping in mind
that ref.~\cite{BA} does not use canonically normalized vector
fields): $c_1 = \sqrt 2$ and $c_4 = 1/\sqrt 6$, giving the result
quoted in the main text:
\eqa \d \cA_m &=& \sqrt2 (\ol\eta_m \gl \veps) + {2 \over \sqrt6}
 (\zbar \g_m \gl \veps) - {2\over \sqrt{6} m}
\pl_m \Bigl( \zbar \gl \veps \Bigr), \nn\\
\gl \d \zeta &=& {1 \over \sqrt6} \cA^*_{mn} \; \g^{mn} \gl \veps
-\, {2 m \over \sqrt6} \, \cA_m \, \g^m \gr \veps, \\
\gl \d \eta_m &=& {1 \over m} \pl_m \left({\sqrt 2 \over 6}
\cA_{ab}^* \g^{ab} \gl \veps + {2 \sqrt{2} m\over 3}
 \cA_a \g^a \gr \veps \right) -\, {2 \sqrt2 \over 3}
 \cA_{mn} \g^n \gr \veps \nn\\
&& + {i \sqrt2 \over 6} \eps_{mnab} \cA^{ab} \g^n \gr \veps -\,
{\sqrt{2} m\over 3} \cA^{n*} \g_{mn} \gl \veps + {2\sqrt{2} m\over
3} \cA^*_m \gl \veps . \nn\eeqa

\subsection{The Massive Spin-1 Multiplet}
As for the spin-3/2 multiplet described above, the global
supersymmetry transformations for the spin-1 multiplet may be
obtained by writing down the most general ansatz which is
consistent with Lorentz invariance, dimensional analysis and a
$U(1)$ symmetry. In this case the $U(1)$ charge is carried by the
two fermions, so we suppose the combination $\beta + i \alpha$
rotates by $e^{i\omega}$ (which is the way we assume $\gr \veps$
also rotates), so $\beta - i \alpha$ and $\gl \veps$ rotate
oppositely. The most general transformation becomes:
\eqa
\d \varphi &=& f_1 \, \ol\veps \gl(\beta + i \alpha) + c.c. \nn\\
\gl \d \beta &=& (\g^m \gr \veps)[f_2 \partial_m \varphi + f_3 v
A_m]
+ f_4v \varphi (\gl \veps) + f_5 A_{mn} (\g^{mn} \gl \veps) \nn\\
\gl \d \alpha &=& -i(\g^m \gr \veps)
[f_2 \partial_m \varphi + f_3 v A_m] \\
&&\qquad\qquad+i f_4v \varphi (\gl \veps) + if_5 A_{mn} (\g^{mn} \gl \veps) \nn\\
\d A_m &=& \partial_m \left[ {f_6 \over v}\, \ol\veps \gl(\beta +
i \alpha) \right] + f_7 \,\ol\veps \g_m \gl (\beta - i \alpha) +
c.c.  \nn\eeqa

Invariance of the action (up to total derivatives) then implies
the relations
\eq f_1 = f_2^* = - f_4, \quad f_3 = f_6^* = f_7, \quad f_7 = -2
\, f_5^*, \eeq
so we can conventionally choose $f_1$ and $f_3$ to arrange that
$\varphi$ has the same transformation rules as does the imaginary
part of the scalar in a chiral multiplet. The required choice is
$f_1 = f_2 = - f_4 = 1/\sqrt{2}$ and $f_3 = 2\, f_5 = - f_6 = f_7
= i/\sqrt{2}$, giving:
\eqa
\d \varphi &=& {1\over \sqrt2} \,\ol\veps \gl(\beta + i \alpha) + c.c. \nn\\
\gl \d \beta &=&{1\over \sqrt2} (\g^m \gr \veps) [\partial_m
\varphi + i v A_m] - {v\over \sqrt2} \,\varphi (\gl \veps) +
{i \over 2\sqrt2}\, A_{mn} (\g^{mn} \gl \veps) \nn\\
\gl \d \alpha &=& -\, {i\over \sqrt2} (\g^m \gr \veps)
[\partial_m \varphi + i v A_m] \\
&&\qquad\qquad-\, {iv\over \sqrt2} \,\varphi (\gl \veps) - \,
{1\over 2\sqrt2} \,
 A_{mn} (\g^{mn} \gl \veps) \nn\\
\d A_m &=& - \partial_m \left[ {i \over v\sqrt2}\, \ol\veps
\gl(\beta + i \alpha) \right] + {i\over \sqrt2} \ol\veps \g_m \gl
(\beta - i \alpha) + c.c. \nn \eeqa
This is the form quoted in the main text.
\end{appendix}


\end{document}